\documentclass[journal]{IEEEtran}

\ifCLASSINFOpdf
\else
   \usepackage[dvips]{graphicx}
\fi
\usepackage{url}

\usepackage[T1]{fontenc}
\usepackage[utf8]{inputenc}
\usepackage[english]{babel}
\usepackage{graphicx}
\usepackage{multirow}
\usepackage{wrapfig}
\usepackage{array}
\usepackage{amsmath}
\usepackage{cases}
\usepackage{tikz}

\makeatletter
\newcommand\footnoteref[1]{\protected@xdef\@thefnmark{\ref{#1}}\@footnotemark}
\makeatother

\begin{document}

\title{F0-based Gammatone Filtering for Intelligibility Gain of Acoustic Noisy Signals}

\author{A. Queiroz, \IEEEmembership{Student Member, IEEE}, and R. Coelho, \IEEEmembership{Senior Member, IEEE}

\thanks{The authors are with the Laboratory of Acoustic Signal Processing (lasp.ime.eb.br), Military Institute of Engineering (IME), Rio de Janeiro, Brazil (email: coelho@ime.eb.br). 
This work was partially suported by the National Council for Scientific and Technological Development (CNPq) 307866/2015 and Fundação de Amparo à Pesquisa do Estado do Rio de Janeiro (FAPERJ) 203075/2016. 
This work is also supported by the Coordenação de Aperfeiçoamento de Pessoal de Nível Superior - Brasil (CAPES) - Grant Code 001.}}

\maketitle

\begin{abstract}

This letter proposes a time-domain method to improve speech intelligibility in noisy scenarios. In the proposed approach, 
a series of Gammatone filters are adopted to detect the harmonic components of speech. 
The filters outputs are amplified to emphasize the first harmonics, reducing the masking effects of acoustic noises. 
The proposed GTF$_\text{F0}$ solution and two baseline techniques are examined considering four background noises with different non-stationarity degrees. 
Three intelligibility measures (ESTOI, ESII and ASII${_\text{ST}}$) are adopted for objective evaluation. 
The experiments results show that the proposed scheme leads to expressive speech intelligibility gain when compared to the competing approaches. 
Furthermore, the PESQ and WSS objective scores demonstrate that the proposed technique also provides interesting quality improvement.

\end{abstract}


\IEEEpeerreviewmaketitle

\section{Introduction}

\IEEEPARstart{A}{coustic} noise masking effects of speech signals 
is still a key element for intelligibility improvement research.
This issue
underlies many applications such as speech syntesis, source localization, and speech and speaker recognition. 
The reduction of noise distortion is a major challenge to improve quality and intelligibility of speech signals. 
Speech enhancement methods have been proposed to treat non-stationary acoustic noises \cite{UMMSE_2012,TAVARES_2016,EMD_ZAO_2014}, 
leading to expressive quality results. 
However, the harmonic components of speech, such as fundamental frequency (F0) and formants, are generally not considered in such solutions. 
F0 estimation is an essential benefit for speech audition, particularly in noisy environment.
Thus, it is here considered as a potential factor to achieve intelligibility gain.

Recently, time-domain adaptive solutions have been designed to deal with the harmonics of the speech signal to reduce the noise effects.  
In \cite{Gael_2017}, the formant center frequencies from voiced segments of speech are shifted away from the region of noise.
This formant shifting procedure \cite{Gael_2016_icassp} simulates the human strategy to provide a more audible signal in noisy environment, 
i.e., the Lombard effect \cite{Lombard_1911}.
Results showed that the Smoothed Shifting of Formants for Voiced segments (SSFV) is able to improve the intelligibility of speech signals in car noise environment. 
A different approach was proposed in \cite{APES_2016}, where linear harmonic models are applied to represent the voiced segments as a sum of sinusoids.
Each voiced frame is reconstructed as a sum of harmonics whose frequencies correspond to the speech F0 and its first integer multiples.
The amplitude and phase estimation filter \cite{APES_Stoica_1999} was applied with the harmonic models (APES$_\text{HARM}$) and led 
to improved signal-to-noise ratios (SNR) of the reconstructed speech signals \cite{APES_2016}.

This letter proposes a new time-domain approach namely GTF$_\text{F0}$ to attain intelligibility gain for speech signals corrupted by acoustic noises. 
In this solution, Gammatone filters are applied to decompose the voiced segments of speech into a series of the harmonics components
with center frequencies defined by integer multiples of F0.
The F0 values are estimated directly from the target noisy speech signal using the HHT-Amp method \cite{zao18}.
The filters outputs are amplified by a gain factor, which emphasizes the first harmonics of the speech signal leading to intelligibility improvement.
In the proposed GTF$_\text{F0}$, the F0 values are not modified since such change would not contribute to an improved speech intelligibility \cite{COOKE_2009}.
Furthermore, it requires no prior knowledge of the speech or noise statistics, which makes GTF$_\text{F0}$ suitable to any kind of noisy environment.

Extensive experiments are conducted to evaluate the proposed scheme for speech intelligibility and quality improvement. 
For this purpose, four acoustic noises with different non-stationarity degrees are used to corrupt the speech signals
considering SNR between -5 dB and 5 dB.
The formant shifting approach (SSFV) \cite{Gael_2017} and the technique based on harmonic models (APES$_\text{HARM}$) \cite{APES_2016} are adopted as baseline.
Three objective intelligibility measures are used to compare the proposed and baseline techniques: ESTOI \cite{ESTOI_2016}, ESII \cite{ESII_2005} and ASII${_\text{ST}}$ \cite{ASII_2015}. 
PESQ \cite{PESQ_2001}, LLR \cite{LLR_1989} and WSS \cite{WSS_1982} are selected to examine the speech quality. 
Results show that the proposed solution outperforms the competing methods in terms of speech intelligibility and quality scores.

\section{F0 Estimation in Non-Stationary Noisy Scenario}

In urban environments, speech signals are usually distorted by acoustic background noises.
Particularly, the F0 estimation accuracy can be highly affected by the presence of acoustic noises. 
This task may become even more challenging when the background noise is non-stationary \cite{zao18}.

\subsection{Non-Stationarity of Noisy Speech Signals}

The non-stationarity degrees of speech signals corrupted by acoustic noises are here examined according to the Index of Non-Stationarity (INS) \cite{Flandrin_10}. 
The INS objectively compares the target signal with stationary references called surrogates.
For each window length $T_h$, a threshold $\gamma$ is defined for the stationarity assumption considering a confidence degree of $95\%$. Thus,
\begin{equation} \label{medins}
\text{INS} \begin{cases}
 \leq \gamma, &  \text{signal is stationary};\\
> \gamma, &  \text{signal is non-stationary}.
\end{cases}
\end{equation}

\begin{figure}[t]
 \centering
 
 \includegraphics[width=\columnwidth]{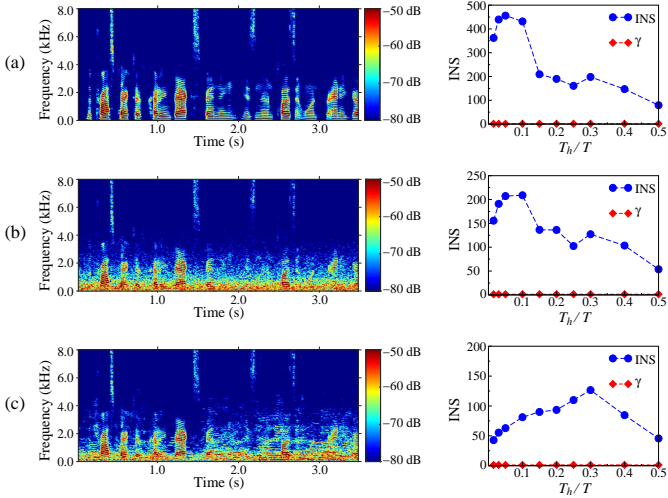}
 \vspace{-0.6cm}
 \caption{Spectrogram and respective INS for (a) clean speech signal and two noisy versions with (b) SSN and (c) Babble noises with SNR of 0 dB.}
 \vspace{-0.4cm}
 \label{INS}
\end{figure}

Fig. \ref{INS} illustrates the spectrogram and INS values obtained for a clean and two noisy versions of the same speech signal.
The INS is computed for different time scales $T_h/T$, where $T$ refers to the total duration of the analyzed signal.
A speech shaped noise (SSN) \cite{DEMAND_2013} and a Babble noise \cite{RSG_1988} are used to corrupt the speech signal with SNR of 0 dB.
Note that the presence of acoustic noises significantly changes the temporal and spectral structures of the speech signal. 
These masking effects can modify the signal harmonic components (F0 and formants). And thus, it may induce speech intelligibility or quality degradation. 
Furthermore, the noise corruption considerably attenuates the non-stationary behavior of the clean speech signal.
For instance, the maximum INS value changes from 450 with clean speech to around 150 when corrupted by the Babble noise.

\vspace{-0.2cm}

\subsection{HHT-Amp F0 Estimation}

The HHT-Amp method applies the Hilbert-Huang transform (HHT) \cite{huang_98} to analyze the target speech signal.
Instead of using the instantaneous frequencies as in \cite{huang06,tao10}, 
the F0 is estimated from the instantaneous amplitude functions of the target signal.
Let $x(t)$ denote a speech signal divided into $Q$ short-time frames $x_q(t), q = 1, 2, \ldots, Q$.
The HHT-Amp method is summarized as follows:

\begin{enumerate}
\item Apply the ensemble empirical mode decomposition (EEMD) \cite{huang09} 
to decompose the sample sequence $x_q(t)$ into a series of intrinsic mode functions (IMF) and a residual $r_q(t)$, 
$x_q(t) = \sum_{m=1}^M \mbox{IMF}_{m,q}(t) + r_q(t)$.

\item Compute the instantaneous amplitude functions as $A_{m,q}(t) = |Z_{m,q}(t)|, m = 1, \ldots, M,$ where the analytic signals are
defined as $Z_{m,q}(t) = \mbox{IMF}_{m,q}(t) + j \, H\{\mbox{IMF}_{m,q}(t)\}$,
and $H\{\mbox{IMF}_{m,q}(t)\}$ refers to the Hilbert transform of $\mbox{IMF}_{m,q}(t)$.

\item Calculate the ACF $r_{m,q}(\tau) = \sum_t  A_m(t) \, A_m(t+\tau)$ of the amplitude functions $A_{m,q}(t), m = 1, \ldots, M$.

\item For each decomposition mode $m$, let $\tau_0$ be the lowest $\tau$ value that correspond to an ACF peak, subject to 
$\tau_{min} \leq \tau_0 \leq \tau_{max}$. The restriction is applied according to the range $[F_{min}, F_{max}]$
of possible F0 values. The $m$-th pitch candidate is defined as $\tau_0 / {f_s}$, 
where $f_s$ refers to the sampling rate.

\item Apply the decision criterion defined in \cite{zao18} to select the best pitch candidate $\hat T_0$. The estimated F0 is
given by $\hat {\text{F0}} = 1 / \hat T_0$.

\end{enumerate}

In \cite{zao18}, it was shown that the HHT-Amp method achieves interesting results in estimating the fundamental frequency of noisy
speech signals. The HHT-Amp was evaluated in a wide range of noisy scenarios, including five acoustic noises with
different non-stationarity degrees. It outperformed four competing estimators in terms of gross error (GE) and mean absolute error (MAE).

\section{Proposed Gammatone Filter Method: GTF$_\text{F0}$}
\label{sec:proposal}

\begin{figure}[t!]
\centering 
\includegraphics[width=1\columnwidth, height=3cm,clip=true,trim=0pt 0pt 0pt 0pt]{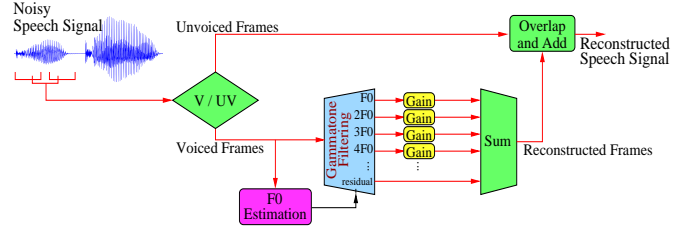}
\caption{Block diagram of the proposed Gammatone Filter method for speech intelligibility gain.}
\vspace{-0.4cm}
\label{fig:schematic}
\end{figure}

The block diagram of the proposed GTF$_\text{F0}$ method is illustrated in Fig. \ref{fig:schematic}.
The target noisy signal $x(t)$ is first split into $Q$ overlapping short-time frames $x_q(t), q = 1, 2, \ldots, Q$, with $50\%$ overlapping. 
Two disjoint sets are formed after the separation of voiced and unvoiced (V/UV) segments. 
$S_v$ is composed by frames that contain voiced speech, and $S_u$ consists of the remaining 
segments, i.e., unvoiced speech and noise. 
For each voiced frame $q \in S_v$, the HHT-Amp method \cite{zao18} is applied to estimate the F0 value from $x_q(t)$.
A total of $L$ Gammatone filters, with center frequencies set to $\hat {\text{F0}}, 2 \,\hat {\text{F0}}, \ldots, L \,\hat {\text{F0}}$, are used to filter the sample sequence $x_q(t)$.
Gain factors are employed to amplify the filters outputs before the reconstruction of the speech frame $\hat x_q(t)$.
Finally, the overlap and add method is applied to all frames to achieve the reconstructed version $\hat x(t)$ of the target speech signal. 

\vspace{-.0cm}
\subsection{Gammatone Filtering}

The Gammatone filter was introduced in \cite{gammatone_orig} to describe the impulse response of the auditory system. 
The time-domain impulse response of the Gammatone filter is defined as 

\begin{equation}
g(t) = a t^{n-1} \cos(2 \pi f_c t + \phi) e^{-2 \pi b t}\, , \, t \geq 0\, ,
\label{eq:gamma}
\end{equation}
where $a$ is the amplitude, $n$ is the filter order, $f_c$ is the center frequency, $\phi$ is the phase, and $b$ is the bandwidth.
In \cite{pat92}, it was shown that a set of fourth-order Gammatone filters are able to represent the magnitude characteristic of the human auditory system.
In the Gammatone auditory filterbank, the bandwidth $b$ presented in \eqref{eq:gamma} is similar to the Equivalent Rectangular Bandwidth (ERB) derived in \cite{pat86},
i.e., $b = 1.019 \, \text{ERB}$.

In the proposed GTF$_\text{F0}$ method, a set of $L$ Gammatone filters $\left\{ h_k(t), k=1 \ldots, L \right\}$ are applied to successively 
filter the input sample sequence $x_q(t)$. 
Each filter $h_k(t)$ is implemented\footnote{Code available at http://staffwww.dcs.shef.ac.uk/people/N.Ma/} considering order $n = 4$,
center frequency $f_c = k \, \hat {\text{F0}}$, and bandwidth $b = 0.25 \, \hat {\text{F0}}$.
In order to align the impulse response functions, phase compensation is applied to all filters, which correspond
to the non-causal filters
\vspace{-0.1cm}
\begin{equation}
h_k(t) = a (t+t_c)^{n-1} \cos(2 \pi f_c t) e^{-2 \pi b (t+t_c)}\, , \, t \geq -t_c\, ,
\label{eq:gamma_c}
\end{equation}
where $t_c = \frac{n-1}{2 \pi b}$ ensures that peaks of all filters occur at $t = 0$.

Let $x_q^0(t) = x_q(t)$, the filtered signals $y_q^k(t), k = 1, \ldots, L$, are recursively computed by
\begin{equation}
\left\{\begin{array}{l}
y_q^k(t) = x_q^{k-1}(t) * h_k(t) \vspace{0.2cm}\\
x_q^k(t) = x_q^{k-1}(t) - y_q^k(t)
\end{array} \right. ,
\quad k = 1, \ldots, L\, .
\label{eq:filters}
\end{equation}
The residual signal is defined as $r_q(t) = x_q^L(t)$ to guarantee the completeness of the input sequence. It means that 
$x_q(t) = \sum_{k=1}^L y_q^k(t) + r_q(t)$.

\begin{figure}[t!]
\centering 
\includegraphics[width=1\columnwidth]{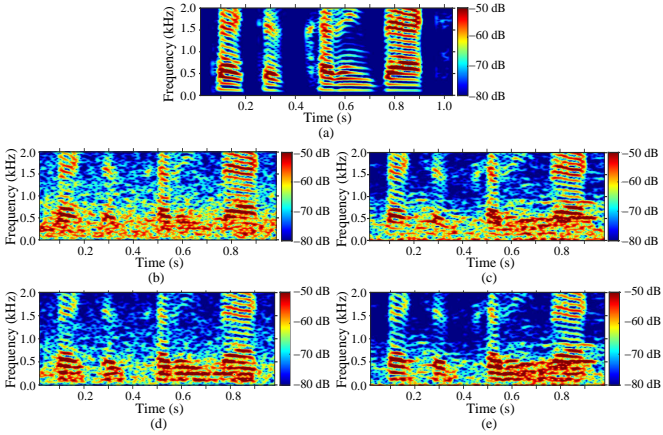}
\vspace{-0.4cm}
\caption{Spectrogram of (a) a clean speech segment, the same signal corrupted with (b) SSN and (c) Cafeteria noise with SNR of 0 dB,
and (d-e) the corresponding signals processed with the proposed GTF$_\text{F0}$ method.}
\label{fig:spectrogram}
\vspace{-0.4cm}
\end{figure}

\subsection{Speech Signal Reconstruction}

After the Gammatone filtering, the amplitude of the output samples $y_q^k(t), k = 1, \ldots, L$, are amplified by a factor $G_k \geq 1$.
The idea is to emphasize the presence of the first harmonics of the fundamental frequency. This will induce speech intelligibility improvement
without introducing any noticeable distortion to the speech signal.
The reconstruction of the voiced frame $q \in S_v$ leads to the sample sequence
\begin{equation}
\hat x_q(t) = \left[ \sum_{k=1}^L G_k \, y_q^k(t) \right] + r_q(t)\, .
\label{eq:frame_rec}
\end{equation}

For the reconstruction of the entire speech signal, the voiced frames obtained in \eqref{eq:frame_rec} and all the remaining frames in $S_u$ are joined together keeping the original frames indices.
Thus, all frames are overlap and added to reconstruct the modified version $\hat x(t)$ of the target speech signal.
The completeness and continuity of $\hat x(t)$ is guaranteed by the adoption of the Hanning window that multiply all frames before the overlap and add method.
This means that the reconstructed signal $\hat x(t)$ and the original signal $x(t)$ would be exactly the same if 
$G_k = 1$ for every $k \in \left\{1, \ldots, L\right\}$.

\begin{table*}[t!] \caption{\label{estoiesii} ESTOI, ESII, and ASII$_{\text{ST}}$ measures [\%] for UNP speech signals}
\vspace{-0.4cm}
\begin{center}
{
\centering
\begin{tabular}{lccccccccccccccccc}
\hline
\multirow{2}{*}&\multicolumn{5}{c}{ESTOI}&&\multicolumn{5}{c}{ESII}&&\multicolumn{5}{c}{ASII$_{\text{ST}}$}\\
\cline{2-6}\cline{8-12}\cline{14-18}

SNR (dB)&-5&-3&0&3&5&&-5&-3&0&3&5&&-5&-3&0&3&5\\\hline
Babble&	0.28&0.33&0.40&0.48&0.53&&	0.34&0.38&0.44&0.50&0.54&&	0.38&0.40&0.45&0.50&0.54 \\

Cafeteria&	0.30&0.35&0.43&0.51&0.57&&	0.36&0.39&0.45&0.52&0.56&&	0.39&0.41&0.46&0.51&0.55\\

SSN&		0.28&0.33&0.40&0.47&0.53&&	0.31&0.34&0.40&0.46&0.50&&	0.35&0.37&0.42&0.47&0.50\\

Volvo&		0.71&0.74&0.79&0.83&0.86&&	0.82&0.85&0.89&0.92&0.94&&	0.77&0.80&0.84&0.87&0.89\\\hline

\end{tabular}
}
\end{center}
\vspace{-0.4cm}
\end{table*}

Fig. \ref{fig:spectrogram} illustrates an example application of the proposed GTF$_\text{F0}$ to a speech signal selected from the TIMIT database \cite{TIMIT_1993}. 
The spectrogram of a clean speech segment and two noisy versions are depicted in Figs. \ref{fig:spectrogram}(a-c).
The corrupted signals are obtained with the SSN and Cafeteria\footnote{\label{note1}Available at www.freesound.org.} noises considering SNR of 0 dB.
It can be noted that the presence of the acoustic noises clearly induce the F0 harmonics to blur, especially the first and second ones.
The GTF$_\text{F0}$ method considering fixed gain of 3 dB to the first $L = 5$ harmonics is applied to these noisy signals.
The resulting spectrograms are shown in Figs. \ref{fig:spectrogram}(d-e).
Note that for both noises the GTF$_\text{F0}$ method achieves more clearly distinguished harmonics when compared to the noisy signals.
This effect may reduce the impact of the acoustic noise to speech intelligibility.

\begin{table*}[t] 
\begin{center} 
\caption{PESQ objective scores for noisy conditions at different SNRs}
\vspace{-0.25cm}
{
\renewcommand{\arraystretch}{1.}
\setlength{\tabcolsep}{3.pt}
\begin{tabular}{lccccccccccccccccccccccccc}
\hline
\multirow{2}{*}&\multicolumn{5}{c}{Babble}&&\multicolumn{5}{c}{Cafeteria}&&\multicolumn{5}{c}{SSN}&&\multicolumn{5}{c}{Volvo}&&Overall\\
\cline{2-6}\cline{8-12}\cline{14-18}\cline{20-24}

SNR (dB)&-5&-3&0&3&5&&-5&-3&0&3&5&&-5&-3&0&3&5&&-5&-3&0&3&5&&Average\\\hline 
UNP&1.98&2.14&2.41&2.71&2.90&&2.15&2.33&2.59&2.89&3.05&&1.91&2.07&2.34&2.64&2.84&&3.75&3.89&\textbf{4.08}&\textbf{4.25}&\textbf{4.35}&&2.86 \\

GTF$_\text{F0}$&\textbf{2.17}&\textbf{2.36}&\textbf{2.66}&\textbf{2.94}&\textbf{3.12}&&\textbf{2.39}&\textbf{2.58}&\textbf{2.86}&\textbf{3.13}&\textbf{3.30}&&\textbf{2.10}&\textbf{2.30}&\textbf{2.61}&\textbf{2.89}&\textbf{3.08}&&\textbf{3.83}&\textbf{3.93}&4.06&4.17&4.23&&\textbf{3.04}\\

SSFV&1.98&2.14&2.42&2.71&2.90&&2.17&2.33&2.59&2.87&3.05&&1.93&2.08&2.35&2.64&2.84&&3.73&3.87&4.05&4.22&4.31&&2.86\\

APES$_\text{HARM}$&2.01&2.18&2.47&2.75&2.91&&2.17&2.35&2.62&2.89&3.05&&1.95&2.14&2.44&2.72&2.90&&3.36&3.47&3.64&3.77&3.84&&2.78\\\hline 

\end{tabular}\label{TABLE_PESQ}
}
\end{center} 
\vspace{-0.6cm}
\end{table*}

\section{Experiments and Results}

Several evaluation experiments are conducted with a subset of the TIMIT speech database \cite{TIMIT_1993}. This is composed of 192 speech signals sampled at 16 kHz, spoken by 24 speakers (16 male and 8 female). 
Each speech segment has an average duration of 3 s. 
Four acoustic noises are applied for the speech signals corruption. The SSN and Cafeteria noises are selected from the DEMAND \cite{DEMAND_2013} and Freesound.org\footnoteref{note1} databases, respectively.
Moreover, Babble and Volvo noises are collected from the RSG-10 \cite{RSG_1988} database. 

The proposed GTF$_\text{F0}$ is implemented considering frames of 32 ms and Gammatone filters bandwidth $b = 0.25 \, \hat {\text{F0}}$.
The first $L = 4$ harmonics are amplified considering the following gain factors: $G_1 = G_2 = 5.0$ dB, $G_3 = 4.0$ dB, and $G_4 = 2.5$ dB. 
The baseline formant shifting approach (SSFV) considers the formant modification function that led to the best results in \cite{Gael_2016_icassp}.
The harmonic models solution with the APES filter (APES$_\text{HARM}$) is applied as described in \cite{APES_2016}.

\vspace{-0.2cm}
\subsection{Objective Intelligibility Evaluation}
\begin{figure}[t]
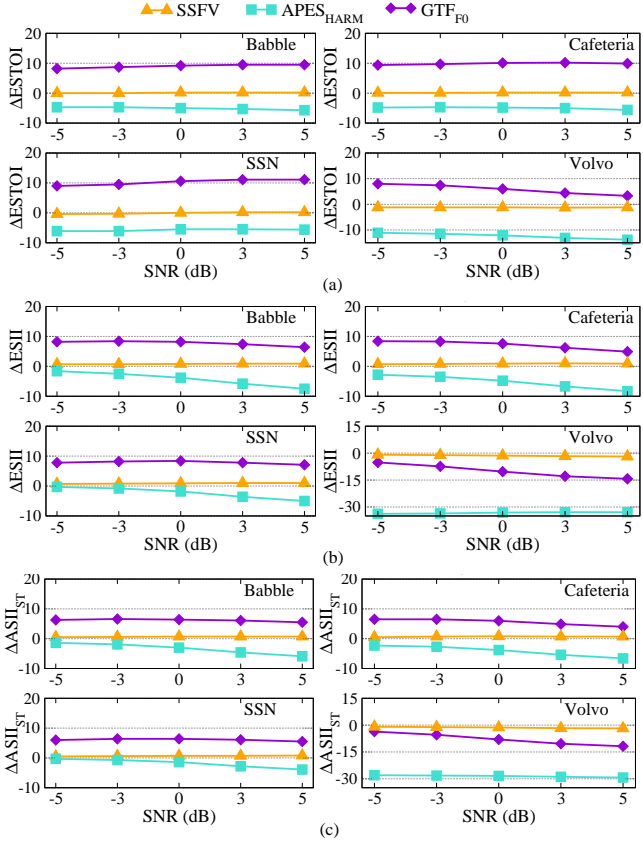

 \centering
  \includegraphics[width=0.95\columnwidth,clip=true,trim=0pt 0pt 0pt 0pt]{ganho_estoi_zao_v2.eps}\\
\vspace{0.1cm}
 \includegraphics[width=0.95\columnwidth,clip=true,trim=0pt 1pt 0pt 32pt]{ganho_esii.eps}\\
\vspace{0.1cm}
 \includegraphics[width=0.95\columnwidth,clip=true,trim=0pt 2pt 0pt 32pt]{ganho_asii.eps}
 \vspace{-0.2cm}
 \caption{(a) $\Delta$ESTOI, (b) $\Delta$ESII, and (c) $\Delta$ASII$_{\text{ST}}$ intelligibility improvement [${\times10^{-2}}$] in four noisy conditions.}
 \label{ganho_estoi}
\end{figure}

Tab. \ref{estoiesii} presents the average ESTOI, ESII and ASII$_{\text{ST}}$ scores obtained with the noisy unprocessed (UNP) speech signals.
The intelligibility improvement achieved with the proposed and baseline solutions are depicted in Fig. \ref{ganho_estoi}.
Note from the ESTOI results that the GTF$_\text{F0}$ leads to the highest gain for all noisy scenarios. 
In average, it outperforms the SSFV approach in 10\% for the Babble, Cafeteria and SSN noises.
For the highly non-stationary Cafeteria noise, the proposed method attains an improvement of 10.1 at 0 dB, compared to 0.4 and -4.8 for the SSFV and APES$_\text{HARM}$ techniques, respectively. 

In terms of ESII and ASII$_{\text{ST}}$ scores, it can be seen that the GTF$_\text{F0}$ leads to the best results for three noise sources: Babble, Cafeteria and SSN. 
The only scenario where this solution does not achieve the highest rates is the Volvo noise. 
In this case, all approaches lead to negative intelligibility gain. 
It is due to the fact that the ESII and ASII$_{\text{ST}}$ scores for Volvo are higher than 0.77 for the noisy signals (refer to Tab. \ref{estoiesii}).
The values are defined as very good intelligibility \cite{intell_1997,intell_2006}. 
Among all the scenarios, GTF$_\text{F0}$ accomplishes the highest overall $\Delta$ESII and $\Delta$ASII$_{\text{ST}}$ of 8.4 and 6.6, respectively, for the non-stationary Babble noise with SNR of -3 dB. 
The APES$_\text{HARM}$ baseline method is outperformed by GTF$_\text{F0}$ and SSFV in all scenarios.

\subsection{Objective Quality Evaluation}

The predicted quality scores computed with PESQ \cite{PESQ_2001} are shown in Table \ref{TABLE_PESQ}. 
As it can be seen, GTF$_\text{F0}$ attains the best PESQ results for three background noise sources: Babble, Cafeteria and SSN. 
Considering the Volvo noise, the unprocessed speech signals present good quality. It means that the highest PESQ scores are obtained by UNP with SNR $\geq$ 0 dB. 
The GTF$_\text{F0}$ attains the best average PESQ value of 3.06, which is 0.17 greater than the noisy signals result.

\begin{figure}[t]
 \centering
 \includegraphics[width=8 cm,height=5.5cm,clip=true,trim=0pt 5pt 19pt 0pt]{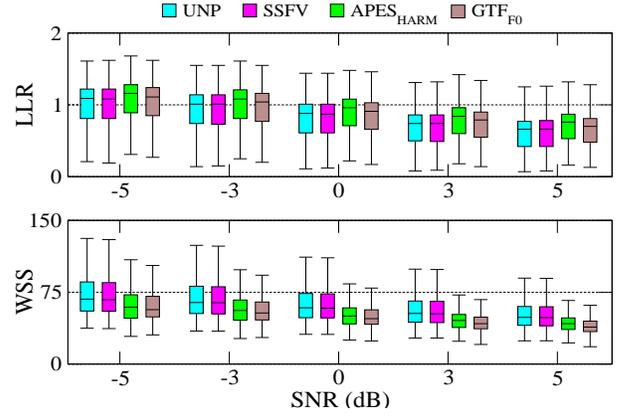} 
 \caption{Mean LLR (top) and WSS (bottom) quality scores for Babble, Cafeteria, SSN and Volvo noises.}
 \label{llrwss}
\end{figure}

The LLR \cite{LLR_1989} and WSS \cite{WSS_1982} measures are also adopted here to objectively examine the speech signal in terms of quality. 
LLR scores are limited in the range $[0,2]$, and just like WSS, smaller values indicate better quality. Fig. \ref{llrwss} shows the results as mean scores computed for the four noise sources at each SNR. 
Note that the GTF$_\text{F0}$ presents the smallest WSS values for all SNR values. These results reinforce the capacity of the proposed solution to emphasize the harmonic components of speech signals, providing improvement
in terms of both intelligibility and quality.

\section{Conclusion}

This letter introduced the time-domain GTF$_\text{F0}$ method to improve intelligibility and quality of speech signals. 
In this solution, F0 estimation and Gammatone filtering are applied to emphasize the first harmonics of the noisy speech signal. 
Four acoustic noises were considered to compose the evaluation scenario. 
Six objective prediction measures were applied to examine the proposed and competitive solutions.
Results showed that GTF$_\text{F0}$ achieved the best intelligibility and quality scores considering ESTOI and PESQ prediction measures for all acoustic noises.


\end{document}